\documentclass[preprint,aps,amsmath,amssymb,floatfix,pdftex]{revtex4}
\usepackage{graphics,color}

\usepackage[dvips]{graphicx}

\begin{document}

\title{Sequence-dependent folding landscapes of adenine riboswitch aptamers}

\author{Jong-Chin Lin$^1$,
Changbong Hyeon$^2$ and
D. Thirumalai$^{1}$ \\
{\it $^1$Department of Chemistry and Biochemistry, Biophysics Program, 
Institute for Physical Science and Technology,
University of Maryland,
College Park, MD 20742} \\
{\it $^2$School of Computational Sciences, Korea Institute for Advanced Study, Seoul 130-722, Korea}
}

\date{\today}


\begin{abstract}
Expression of a large fraction of genes in bacteria is controlled by
riboswitches, which are found in the untranslated region of mRNA.
Structurally riboswitches have a conserved aptamer domain to which a
metabolite binds, resulting in a conformational change in the downstream
expression platform. Prediction of the functions of riboswitches requires a
quantitative description of the folding landscape so that the barriers and
time scales for the conformational change in the switching region in the
aptamer can be estimated. Using a combination of all atom molecular dynamics
(MD) and coarse-grained model simulations we studied the response of adenine (A)
binding {\it add} and {\it pbuE} A-riboswitches to mechanical
force. The two riboswitches  contain a structurally
similar three-way junction  formed by three paired helices,
P1, P2, and P3, but carry out different functions. Using pulling simulations,
with structures generated in MD simulations, we show that after P1 rips
the dominant unfolding pathway in {\it add}  A-riboswitch is the
rupture of P2 followed by unraveling of P3. In the {\it pbuE}
 A-riboswitch, after P1 unfolds P3 ruptures ahead of P2.
The order of unfolding of the helices, which is in accord with single molecule
pulling experiments, is determined by the relative stabilities of the
individual helices. 
Our results show that the stability of isolated helices determines the order of assembly and response to force in these non-coding regions. We use the simulated free energy profile for {\it pbuE}
 A-riboswitch to estimate the time scale for allosteric switching, which shows that this riboswitch is under kinetic control lending additional support to the conclusion based on single molecule pulling experiments.
A consequence of the stability hypothesis is that a single point
mutation (U28C) in the P2 helix of the {\it add}  A-riboswitch,
which increases the stability of P2, would make the folding landscapes of the two riboswitches similar. This prediction can be tested in single molecule pulling experiments.
\end{abstract}


\maketitle

Riboswitches, RNA elements located in the untranslated region of mRNAs, 
regulate gene expression by sensing and binding target cellular 
metabolites \cite{Winkler2005}. Function of riboswitches involves allosteric
communication between a conserved aptamer domain and the downstream expression platform. In bacteria, specific metabolites bind to the aptamer domain with 
exquisite selectivity, resulting in a change in the folding patterns of the
expression platform, whose conformation controls transcription
termination or translation initiation \cite{Serganov2004,Serganov2013}. 
Purine riboswitches \cite{Edwards2007}, which are among the simplest, 
display remarkable discrimination in binding metabolites and carry out entirely
different functions despite the structural similarity of the
metabolite-binding aptamer domains (Fig.1). 
Surpringly, even the riboswitches that bind the same 
metabolite function differently in different species 
\cite{Mandal2003NSMB,Mandal2003Cell}. For instance, the {\it pbuE} adenine (A) 
riboswitch activates gene expression upon metabolite binding by disrupting the
formation of a terminator stem in the downstream expression platform. 
The absence of the terminator hairpin upon ligand binding prevents the 
polymerase from engaging with the  poly-U track, resulting in 
completion of transcription \cite{Mandal2003NSMB}.
In contrast, the {\it add} adenine riboswitch 
is a translational activator, which upon ligand binding facilitates the
ribosome to recognize the Shine-Dalgarno sequence, thus initiating
translation \cite{Serganov2004}. 
Thus, we classify {\it pbuE} A-riboswitch as an on-switch riboswitch, 
which implies that gene expression is promoted when the metabolite adenine 
binds. From this perspective, the {\it add}  A-riboswitch is an 
on-switch for translation. 

Purine riboswitch aptamers contain a three-way junction, which is formed by 
helix P1 and hairpins P2 and P3 and are stabilized by tertiary interactions in 
the folded state (Fig. 1). The ability of riboswitches, and more generally 
RNA, to adopt alternate folds, a consequence of the modest stability
gap \cite{Guo1992,Goldstein1992} compared to proteins \cite{Thirum05Biochem},
is crucial in their ability to regulate transcription and translation. 
In both  {\it pbuE} and {\it add} A-riboswitches, part of the aptamer
region (a structural element in blue
located at the  3'-end of P1 in Fig. 1)
is involved in the formation of alternative hairpin structure with 
nucleotides in the downstream expression platform. The time scales and the barriers associated with the  switching strands to  form hairpin with the downstream expression platform determine the dynamic range and efficiency of ribowitch function. Hence, it is important to quantitatively obtain the folding landscapes of the aptamers, which in turn would provide insights into the differences in the functions of the two structurally similar riboswitches. 

Single molecule pulling experiments \cite{Greenleaf2008,Frieda2012} and 
carefully designed computer simulations \cite{Lin2008JACS,Lin12JPCL} are 
ideally suited to obtain the sequence-dependent folding landscapes of riboswitches. 
In the Laser Optical Tweezers (LOT) experiments a constant 
mechanical force ($f$) is applied to the ends of the riboswitches through 
handles, and the response is monitored by measuring the molecular extension, 
$R$, which is conjugate to $f$. Such experiments have been performed on several 
riboswitches \cite{Anthony12PNAS} including {\it pbuE} \cite{Greenleaf2008} 
and {\it add} adenine riboswitches \cite{Neupane2011}. 
The structural changes that occur upon application of force are typically 
characterized using the free energy profiles, $F(R)$,  which provide estimates of the barriers for rupture of various helices. The length gain upon unfolding can be used to decipher the order in which the paired helices unravel. 

In our previous report \cite{Lin2008JACS}, we reported the order of force-induced rupture of {\it add} A-riboswitch using simulations of coarse-grained models. 
The predicted structural transitions in {\it add} A-riboswitch is 
different from the experimentally inferred pathway for {\it pbuE} A-ribsoswitch \cite{Greenleaf2008}
even though they have nearly identical three-dimensional structures (Fig. 1). 
The first event in the predominant unfolding pathway of the aptamers is the 
disruption of helix P1 and the binding pocket formed by the helix junction (Fig. 1). 
It is the subsequent order of unfolding (P2 unfolds before P3
predicted in our study on {\it add} A-riboswitch aptamer) that differs from 
the experimental results for {\it pbuE} A-riboswitch aptamer. 
These results were explained \cite{Lin2008JACS} by noting that the differences in the folding landscapes is due to variations in the stabilities of individual
helices (Fig. 1).
Here, we combine molecular dynamics
and coarse-grained simulations to further explore the differences in the 
folding landscapes of these two structurally related aptamers.
In order to establish the stability hypothesis, as the principle governing assembly of RNA, we first used all atom molecular
dynamics (MD) simulations to obtain putative structures for {\it pbuE}
 A-riboswitches for use in coarse-grained simulations. The combined approaches show
that the dominant unfolding pathway is similar to that inferred from
experiments  \cite{Greenleaf2008}. 
The present work also demonstrates that simulations, at different 
levels of description, can not only predict the outcomes of experiments but 
also yield (qualitative) insights into sequence-dependent 
differences in the response of  even structurally related RNA aptamers to force.


\section*{RESULTS}

\subsection*{Stability hypothesis holds even in the presence of tertiary interactions}
We used the  mfold package \cite{Zuker2003} to estimate the free energy of the 
isolated secondary paired helices in the aptamers (Fig.1).  
The stability of P1 is comparable in both {\it pbuE} and {\it add} 
 A-aptamers.
The P3 helix is more stable by 1.2 kcal/mol in {\it add} A-aptamer
(Fig. 1a), whereas the predicted free energy of the 
secondary structure of P3 is larger than P2 by 2 kcal/mol in 
{\it pbuE} A-riboswitch aptamer (Fig. 1b). 
From stability considerations alone \cite{Lin2008JACS}, we predicted that the order of unfolding
under force from the folded (N) to the globally unfolded state U should be
$N\rightarrow \Delta P_{1} \rightarrow \Delta P_{1} \Delta P_{2} \rightarrow U$
where $\Delta P_{1}$ means rupture of P1 and $\Delta P_{1} \Delta P_{2}$
implies that both P1 and P2 are unfolded. This prediction has subsequently been quantitatively validated in experiments \cite{Neupane2011}. Similarly, 
the predominant unfolding pathway in {\it pbuE} A-riboswitch is
expected to be  
$N\rightarrow \Delta P_{1} \rightarrow \Delta P_{1} \Delta P_{3} \rightarrow U$
(see Fig. 1b).
The theory based on relative stabilities of isolated P2 and P3 readily 
explains the experimental findings. 
However, it is important to examine whether the stability hypothesis is
valid in the presence of tertiary contacts as well. Accordingly,
we used a combination of all atom MD and Brownian dynamics simulations to
determine the $F(R) = -k_BT ln P(R)$ ($P(R)$ is the distribution of the extension, $R$, at a given $f$) profile of the \emph{pbuE}  A-riboswitch, so that a direct comparison with $F(R)$ obtained in simulations and experiments can be made.  

Since the structure of the {\it pbuE} A-riboswitch aptamer is not 
available, we used the crystal structure of {\it add} adenine riboswitch in the
metabolite-bound state \cite{Serganov2004} as a template in all atom MD
simulations to generate putative structures for use in the coarse-grained
self-organized polymer (SOP) model simulations (see Methods). 
We replaced the nucleobases in the {\it add} A-riboswitch with 
those in the {\it pbuE} A-riboswitch. We then used this
structure with the {\it pbuE} sequence as the initial conformation and 
performed all-atom MD simulations for 95 ns to generate putative ensemble of folded 
structures for {\it pbuE} A-riboswitch aptamer at T = 310 K 
(see Methods). The MD-generated structures are then taken as the native 
structures in the coarse-grained simulations in which the RNA is
represented using the SOP model \cite{Hyeon2006}.


After the first 10 ns of molecular dynamics run, 
the root-mean-square deviation (RMSD) of the positions of the backbone nucleotides 
of the {\it pbuE} A-riboswitch aptamer does not change significantly
(Fig. 2a). However, there are instances of larger fluctuations, which suggest that the native state ensemble generated in atomically detailed MD simulations is somewhat heterogeneous.
Both P2 and P3 remain folded during the simulations as indicated by the
stabilizing potential energies of the helices, calculated using the SOP 
energy function with MD snapshots as the native structures.
For all the snapshots recorded at every 10 ps, the average difference in
non-bonded energy between P2 and P3 is about 
$\Delta U_{nb} = U_{nb,P2}-U_{nb,P3} = -3.25$ kcal/mol with a fluctuation in energy, $\sqrt{\langle \delta(\Delta U)^2\rangle}=2.4$ kcal/mol.  
The combined use of MD simulations and SOP representation of the aptamer
shows that P2 is more stable than P3 in {\it pbuE} A-riboswitch.  For the {\it add} A-riboswitch the drift in RMSD (Fig. 2b) is less and the $\Delta U_{nb} U_{nb} = U_{nb,P3}-U_{nb,P2}$ is negative at all times indicating that P3 is more stable than P2.
Thus, we surmise that for both the riboswitches the presence of tertiary interactions does not affect the stability of the paired helices.

\subsection*{Response of {\it pbuE}  A-aptamer to force}

We take snapshots from the MD simulations saved at every 500 ps for $t >$ 10 ns,
which results in a total 168 structures, as the  putative ensemble of native structures  in coarse-grained pulling simulations. The average $\Delta U$ and its 
deviation for these chosen structures are similar to the values for all 
snapshots recorded. Hence, it is reasonable to study the 
stability of the helices using the ensemble of MD generated structures,
assuming that the 
aptamer fluctuates around the average native structure in equilibrium.

The interplay of stability of helices P2 and P3 in modulating the folding
landscape of the aptamers is illustrated by 
investigating the response of helices to mechanical force. 
Among the 168 trajectories generated using the MD 
snapshots  subsequently used in SOP simulations, the folding probability of P2 and P3 varies depending upon the precise starting conformation of the 
native state ensemble. For example, in the trajectory shown in Fig 3a, 
at $f = 13$ pN, both 
helices P2 and P3 hop back and forth between the folded and unfolded states, while
helix P2 spends more time in the folded state than P3. 
This shows that P2 is 
more stable than P3.
The time traces of the molecular extension, $R$, and the free energy landscape
(Fig. 3b and 3c) show three distinguishable folding intermediate states for the {\it pbuE} 
A-riboswitch aptamer at $f = 13$ pN. The aptamer switches 
between unfolded, P2 or P3 folded,
and both P2 and P3 folded (P2/P3) states, with the corresponding extensions 
$R$ around 21.5 nm, 16 nm, and 10 nm, respectively. The intermediate state 
at $R \sim 16$  nm indicates only one helix is folded. 
The probability averaged over time that P2 is folded is $\approx$ 0.90 whereas the probability that P3 is intact is $\approx 0.10$. 
The folding probability of P2 and P3 remaining intact varies when
choosing different MD snapshots as the native structure.

When $f=12$ pN, the riboswitch switches to the folded state, which is the most stable state with a large unfolding free energy barrier ($\approx$ 8 $k_BT$). 
On an average, with the use of 168 different MD snapshots as native structures,
we find that in the intermediate state containing only one folded helix, 60\%
of the time is P2 folded, suggesting P2 is more stable than P3 in 
{\it pbuE} A-riboswitch.  These results are in qualitative agreement with experiments.

We can define the free energy difference between the only-P2-folded
and only-P3-folded state by the ratio,
{\it i.e.}, $\Delta \Delta G = - k_{B}T$ln$(F_{P2}/F_{P3})$, and obtain
the histograms of $\Delta \Delta G$ for all the trajectories 
(Figs. 4a and 4b). On an average the only-P2-folded state is more stable than only-P3-folded state
by about 0.5 $k_{B}T$ (an underestimate arising from potential inaccuracies in the all atom MD force fields) for the {\it pbuE} A-riboswitch aptamer.
At $f = 13$ pN, the relative stability depends modestly 
on the pulling direction; about 45\% of the trajectories show P2 folded more 
than P3 when pulling from the 5'-end, while the percentage becomes 73\% when 
pulling from the 3'-end. Overall, the stability of P2 is larger 
than P3 for the {\it pbuE} riboswitch aptamer, which is in qualitative accord  
with the experimental results \cite{Greenleaf2008}.


\subsubsection*{{\it add} {\bf A-riboswitch aptamer}}
For comparison and to complement our earlier studies based on coarse-grained model \cite{Lin2008JACS}, we also perform MD simulations for the {\it add} 
A-riboswitch aptamer for 75 ns starting with the crystal structure. 
As shown in Fig.  2b, the dynamics of the system becomes stationary after 10 ns. 
We take snapshots at every 500 ps for $t > 10$ ns as the native structures for 
subsequent use in coarse-grained simulations. 
The histograms of the difference in the folding probability between P2 
and P3 for the 131 trajectories each 27 ms long 
also indicates a larger relative stability towards P3
( Figs. 4c and 4d). About three out of 
four trajectories have P3 spending more time folded than P2 with the pulling direction having little effect on the relative stability of the
two helices. 
We find that 
 the only-P3-folded state is more stable than the only-P2-folded 
state by about 1 $k_{B}T$ for the {\it add}  A-riboswitch aptamer. This is opposite
to the {\it pbuE}  A-riboswitch aptamer, where the only-P2-folded state is more
stable than the only-P3-folded state. 
Hence, despite the similar tertiary structures
of {\it pbuE} and {\it add} A-riboswitch aptamers, the relative stabilities
of P2 and P3 found in our simulations are different because of variations in the sequence. 


\section*{CONCLUSIONS}
The differences in the folding landscapes under tension between  \emph{add} and \emph{pbuE} A-riboswitches (both bind purine) were explained based on the stability hypothesis \cite{Lin2008JACS} according to which the order of unfolding is determined by the stability of the individual helices. 
Here, we have further established the validity of this proposal using a
combination of all atom molecular dynamics and coarse-grained
 (CG) simulations. In particular, the multi scale simulations confirm 
that helix P2 is more stable than P3 in {\it pbuE} 
adenine riboswitch aptamer, which is the opposite to that found in {\it add} 
A-riboswitch aptamer. Despite the similarity of the aptamer structures, the sequence difference results in variations in the 
relative stability of helices P2 and P3. 
Surprisingly, the differences in the local contacts within helices 
are enough for our simple model to capture the relative stability of helices 
in {\it add} and {\it pbuE} A-riboswitch aptamers. However, further 
investigations of the conformations of the aptamer coupled with the downstream expression platform 
should be studied to have a complete understanding of the mechanism 
underlying the functions of the purine riboswitches.
We conclude with the following remarks. 

(1) A consequence of the stability hypothesis is that the relative probability of unfolding P3 is (using the free energies in Fig.1a)  \emph{add} A-riboswitch should be $\sim e^2/(1+e^2)\approx 0.9$. From the histogram of $\Delta\Delta G$, calculated using MD generated structures in CG simulation, this probability is $\approx 0.8$, which is comparable to the estimate based on the stability hypothesis. 
A similar calculation based on the free energy given in Fig.1b for \emph{pbuE} A-riboswitch predicts that the probability that P3 folds before P2 is only $\approx 0.04$. 
 Although the multi scale simulations are in qualitative agreement with experiments qualitatively, the combination of MD and CG simulations suggests that this probability is nearly ten times larger. 
  We attribute the discrepancy to plausible deficiencies in the current nucleic acid force fields.  
  Only recently tetraloop (four nucleobases) structures have been accurately predicted by significantly altering the current RNA foce fields \cite{Chen13PNAS}.  Thus, we are only able to obtain qualitative agreement between experiments and simulations for \emph{pbuE}  A-riboswitch, whereas our earlier predictions for \emph{add} A-riboswitch based on CG simulations \cite{Lin2008JACS} agree quantitatively with single molecule pulling experiments \cite{Neupane2011}. It also follows that currently CG model simulations are more accurate than atomically detailed simulations for nucleic acids.

(2) The stability hypothesis for RNA assembly is similar to the ideas used to predict forced-unfolding of proteins \cite{Klimov00PNAS} where  
 it was shown that the order of unfolding of proteins is determined by stability of tertiary interactions associated with a given secondary structural element. 
In both proteins and RNA $f$-dependent landscape is determined by the native topology. 
Because interactions favoring secondary structure formation are much greater than tertiary interactions in RNA, the $f$-dependent landscape is essentially determined by the relative free energies of isolated helices. 
This justifies the stability hypothesis. 

(3) The free energy profile in Fig. 3c could be used to obtain an approximate bound on the time scales in which switching of the region in P1 responsible for transcription control exerted by \emph{pbuE} A-riboswitch. An effective free energy barrier for this switch at $f = 0$ is $\Delta F^{\ddagger}(0) \approx \Delta F^{\ddagger}(f) + f \Delta X^{\ddagger}$. In Fig 3c, $f$= 12pN, $\Delta F^{\ddagger}(f) \approx 8k_BT$, and $\Delta X^{\ddagger} \approx$ 2nm, which gives $\Delta F^{\ddagger}(0) \approx 14 k_BT$. The time scale for switching is $\tau_S \tau_0 exp(\frac{\Delta F^{\ddagger}(0)}{k_BT}$. Using the estimate for the prefactor $\tau_0 \approx 1 \mu s$ \cite{Hyeon12BJ} we obtain $\tau_S \approx$ 1.2 s. Upon binding adenine this time scale is about an order of magnitude greater. Synthesis of downstream nucleotides occur at a rate 20nt/s. Thus, the decision to terminate transcription  must occur in a small window of time on the order of (2-4) seconds (depending on the length of transcript in the expression platform) before metabolite binds. Thus, it is likely that the folded apamer regulating transcription \emph{pbuE} A-riboswitch cannot reach thermodynamic equilibrium as the number of folding transitions in the time window cannot exceed unity. We surmise that the function of \emph{pbuE} A-riboswitch is under kinetic control lending further support to the conclusion reached in single molecule pulling experiments.

(4) Based on the stability hypothesis, we make a prediction for pulling
experiments in a mutant of \emph{add} A-riboswitch. The main reason for the
different energies of P2 between the two purine riboswitches is that there is
one G-U and two G-C base pairs in P2 in  {\it add} A-riboswitch,
whereas there are three G-C base pairs in P2 in
 {\it pbuE} A-riboswitch.
A U28C point mutation in  {\it add} A-riboswitch, resulting in
three G-C base pairs  in P2, would make the secondary free
energy of P2  be -7.3 kcal/mol.
Thus, in the U28C mutant  of {\emph{add} A-riboswitch P2 would be more stable than P3 by about 1.1 kcal/mol.
As a consequence, we predict that the very order of unfolding of \emph{add} A-riboswitch would be reversed. 
The folding landscape of the U28C \emph{add} A-riboswitch would be qualitatively similar to  the WT \emph{pbuE} riboswitch.
\\
 
\section*{METHODS}

Our goal is to predict the structural basis of the free energy landscape 
differences between {\it add} A-riboswitch aptamer and {\it pbuE} A-riboswitch
aptamer. Because the structure of {\it pbuE} A-riboswitch aptamer is 
unavailable, we used the following multi scale computational strategy. 
To create Self-Organized Polymer (SOP) representation of {\it pbuE} A-riboswitch aptamer, we generated an
ensemble of equilibrated structures using all atom molecular dynamics 
simulations using the RNA segment for the {\it pbuE} A-riboswitch aptamer with the 
initial structure corresponding to the {\it add} A-riboswitch aptamer. 
Consistency between MD and coarse-grained simulations allows us to infer the  
robustness of our conclusions.

\subsection*{Self-Organized Polymer (SOP) Model}

To model the riboswitch aptamer, we use a modified form of the self-organized 
polymer (SOP) model \cite{Hyeon2006} that has been used with considerable success in describing 
complex processes ranging from folding \cite{Reddy12PNAS} to allostery in proteins \cite{Hyeon06PNAS} and forced-unfolding of RNA 
\cite{Hyeon2006}. In addition, other studies have also established that coarse-grained models are successful in providing the dynamics and folding of riboswitches \cite{whitford2009nonlocal,feng2011cooperative}.
In the simplest version of the SOP model, 
each nucleotide 
as well as the metabolite adenine is represented as a single interaction site. 
The potential energy of the aptamer in the presence of bound adenine is
\begin{eqnarray}
V_{T} = V_{APT} + V_{APT-AD},
\end{eqnarray}
where the energy functions of the aptamers are given by
\begin{eqnarray}
V_{APT} = V_{FENE} + V_{NB},
\end{eqnarray}
with
\begin{align}
V_{FENE} &= -\sum_{i=1}^{N-1}\frac{k}{2}R_{0}^{2}ln\left( 1-\frac{(r_{i,i+1}-r_{i,i+1}^{0})^{2}}{R_{0}^{2}}\right)
\end{align}
and
\begin{align}
V_{NB} &= \sum_{i=1}^{N-3}\sum_{j=i+3}^{N}\varepsilon_{h}\left[\left(\frac{r_{ij}^{0}}{r_{ij}}\right)^{12} - 2 \left(\frac{r_{ij}^{0}}{r_{ij}}\right)^{6}\right]\Delta_{ij} \nonumber \\
&+ \sum_{i=1}^{N-2}\varepsilon_{l}\left(\frac{\sigma^{*}}{r_{i,i+2}}\right)^{6} \\ 
&+ \sum_{i=1}^{N+3}\sum_{j=i+3}^{N}\varepsilon_{l}\left(\frac{\sigma}{r_{ij}}\right)^{6} (1-\Delta_{ij}) \nonumber 
\end{align}
The term $V_{FENE}$ in Eq. (3) describes the chain connectivity with $k =$ 2000 
kcal/(mol $\times$ nm$^{2}$), $R_{0} =$ 0.2 nm, $r_{i,i+1}$ is the distance 
between two adjacent nucleotides $i$ and $i+1$, and $r_{i,i+1}^{0}$ is the 
distance in the native structure. The non-bonded interaction term, $V_{NB}$, 
in Eq. (4) accounts for the stabilizing forces between the nucleotides that 
are in contact in the native state. The interactions between the nucleotides 
that form non-native contacts are taken to be repulsive. Two nucleotides $i$ 
and $j$ are in native contact with $\Delta_{ij} = 1$ (Eq. (4)) if the distance 
$r_{ij}$ between them in the native structure is within a cutoff distance, 
$R_{c} = 1.3$ nm, for $|i - j| > 2$. If $r_{ij}$ exceeds $R_{c}$, then 
$\Delta_{ij} =$ 0. The interaction between adenine and the aptamer, 
$V_{APT-AD}$, is taken to be, 
\begin{eqnarray}
V_{APT-AD} &=& \sum_{i=1}^{N}\varepsilon_{A}\left[\left(\frac{r_{i,A}^
{0}}{r_{i,A}}\right)^{12} - 2 \left(\frac{r_{i,A}^{0}}{r_{i,A}}\right)^{6}\right]\Delta_{i,A} \nonumber \\
&+& \sum_{i=1}^{N}\varepsilon_{l}\left(\frac{\sigma^{*}}{r_{i,A}}\right)^{6} \\
&+& \sum_{i=1}^{N}\varepsilon_{l}\left(\frac{\sigma}{r_{i,A}}\right)^{6} (1-\Delta_{i,A}) \nonumber
\end{eqnarray}
We set $\varepsilon_{A}$ as the interaction 
between adenine and the nucleotides that are in contact with adenine. 
In the native structure of the {\it add} adenine riboswitch, there are 7 
nucleotides that are in contact with adenine. To prevent adenine  from 
drifting away from the aptamer during the simulations, a restraining potential 
is added between the metabolite and U74.
 
We use two values for the parameter $\varepsilon_{h}$ (see Eq. 3) depending 
upon whether 
the  two nucleotides in native contact are engaged in a secondary or a 
tertiary interaction. If the two nucleotides are within a hairpin or helix, 
$\varepsilon_{h} = \varepsilon_{s}$, otherwise, 
$\varepsilon_{h} = \varepsilon_{t}$. From the largely hierarchical nature of 
RNA folding process \cite{Brion1997}, it follows that the strength of the 
secondary interaction 
is greater than the tertiary interaction. In our simulations, we set 
$\varepsilon_{s} = 0.7$ kcal/mol, and $\varepsilon_{t}/\varepsilon_{s} = 1/2$. 
The strength of the repulsive interaction is taken to be 
$\varepsilon_{l} = 1.4$ kcal/mol for 
non-native contacts. We chose $\sigma = 0.7$ nm, and $\sigma^{*} = 0.35$ nm 
for $i$, $i+2$ pairs to prevent the flattening of the helical structure when 
the overall repulsion is large. Our previous works \cite{Lin2008JACS} have 
shown that riboswitches and other RNA constructs \cite{Hyeon2006,Hyeon2007}
subject to tension are 
accurately described using the chosen range of parameters.

\subsection*{Brownian Dynamics}
The dynamics of the system is described using the Langevin equation in the 
overdamped limit. The equation of motion for the $i^{th}$ nucleotide is
\begin{eqnarray}
\gamma m_{i}\frac{dx_{i}}{dt} = -\frac{\partial V_{i}}{\partial r_{i}} + F_{i}(t),
\end{eqnarray}
where $\gamma$ is the friction coefficient, $m_{i}$ is the mass of nucleotide 
$i$, and $F_{i}(t)$ is the random force, which satisfies
\begin{eqnarray}
\langle F_{i}(t)\rangle = 0,
\end{eqnarray}
and
\begin{eqnarray}
\langle F_{i}(t)F_{i}(t')\rangle = 2 k_{B}T\gamma m_{i}\delta(t-t'),
\end{eqnarray}
where the averages are over an ensemble of realizations or trajectories.

The integration step in the Brownian dynamics  simulations is 
$ \Delta \tau_{H} = \frac{\gamma \varepsilon_{h}}{k_{B}T}h\tau_{L}$, where the 
typical value for $\tau_{L}$ for nucleotides is 4 ps \cite{Hyeon2006}, and 
the integration step size $h = 0.03 \tau_{L}$. For the overdamped limit, we use
$\gamma = 100 \tau_{L}^{-1}$, which approximately corresponds to the friction 
coefficient for a nucleotide in water \cite{Hyeon2007}. For a typical value of
$\varepsilon_{h} = 0.7$ kcal/mol, this results in an integration time step of 
about 14 ps. To unfold the aptamer, an external force is applied to the 5'-end 
of the aptamer, while the 3'-end is fixed.

\subsection*{All-atom Molecular Dynamics (MD) Simulation}

We used MD simulations to obtain approximate native structures for {\it pbuE} 
adenine riboswitch aptamer for use in coarse-grained pulling simulations. 
The NAMD 2.6 molecular dynamics simulation 
package \cite{Phillips2005} and CHARMM force field \cite{MacKerell2001}
were used in all energy minimization.

A total of 71 nucleotides of RNA with the metabolite, adenine, bound and
5 bound magnesium ions were taken from the crystal structure of
the aptamer 
domain of the {\it Vibrio vulnificus add} A-riboswitch (PDB code: 1Y26) 
\cite{Serganov2004}. By exploiting the structural similarity between the two 
riboswitches, we threaded the sequence of {\it pbuE} A-riboswitch through the 
structure of {\it Vibrio vulnificus add} A-riboswitch. We then added 60 sodium
ions, with each placed around the phosphate group of 
RNA backbone, to make the whole system charge neutral.
The system was then 
solvated using the SOLVATE program in the VMD package \cite{Humphrey1996} in an
explicit TIP3P \cite{Jorgensen1983} periodically replicated water solvent box.
A buffer of water around the solute of at least 15 \AA $ $ in all directions
 were added, resulting in total 63,632 atoms in the system. While keeping the positions of RNA, metabolite 
adenine, and magnesium ions fixed, the water and sodium ions were 
allowed to move and the energy is minimized for 2000 cycles. 
Subsequently, the ions and the solvent were relaxed 
by performing molecular dynamics at constant volume, for 600 ps. In the first 
200 ps, the temperature was increased from $T = 0$ K to
$T = 310$ K gradually, and during the second 200 ps,
the temperature remained at $T = 310$ K. 
In the final 200 ps, the temperature was reduced from
$T = 310$ K to 0 K gradually. 

Non-bonded interactions were smoothly switched to zero between 10 and 12 \AA, 
yielding a cutoff radius of 12 \AA. We used 
particle-mesh Ewald algorithm for long-range electrostatic 
interactions with a grid spacing smaller than 1 \AA \cite {Darden1993}. 
The integration time step in MD simulations was 2 fs. We used the SHAKE method 
\cite{Ryckaert1977} for enforcing constraints. The energy of the system
 was then minimized by gradually releasing the 
positional restraint of RNA, the metabolite adenine, and magnesium ions in the 
following way: 1000 energy minimization cycles for each n in the harmonic 
positional restraints of $10^{n/4}$ kcal/(mol$\cdot$ \AA$^{2}$), 
$n$ = 4, 3, 2, 1, 0, -1, …, -15, on RNA, adenine, and magnesium ions. 
We then heated the system from $T = 0$ K to $T = 310$ K for 2 ns without any 
restraint at constant volume, and then kept the system at fixed $T = 310$ K 
for 1 ns. The system was then equilibrated by performing molecular dynamics at 
constant pressure of $p = 1$ atmosphere and constant temperature of $T = 310$ K
for 2 ns with time step being 1 fs. Finally, we performed a 95 ns production 
run at constant $N$, $p$, and $T$ conditions.
The structures for use in the coarse-grained simulations were
obtained from the production run. 
For reasons explained in the final section this procedure is only qualitatively reliable. 
\\

{\bf Acknowledgements:} Part of this work was done while DT was a KIAS scholar. This work was supported in part by a grant from the National Institutes of Health (GML089685) to DT.


\clearpage
\section*{Figure Legends}

\begin{figure}{}
\begin{center}
\centerline{\includegraphics[width=.75\textwidth]{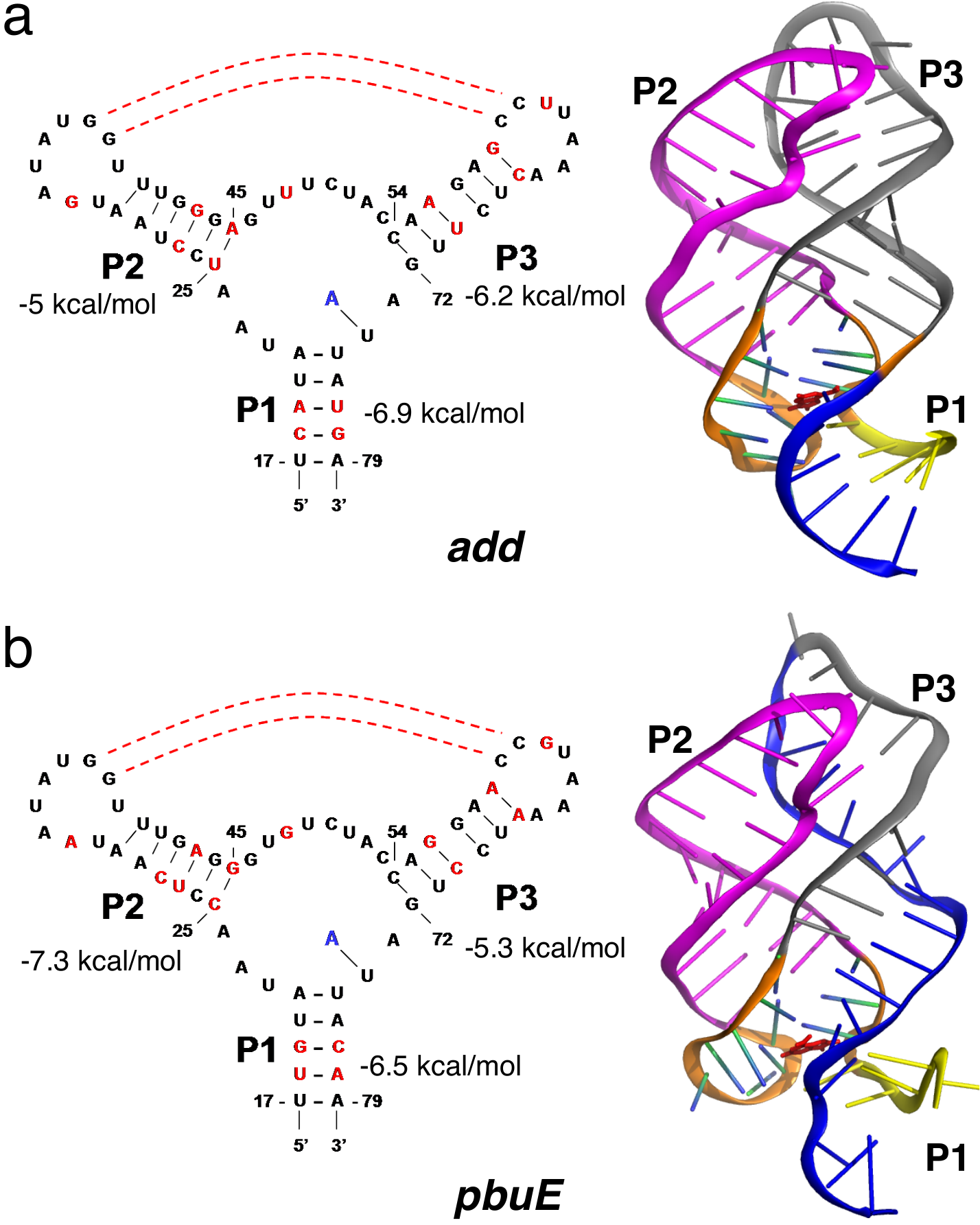}}
\caption{Secondary and tertiray structures of adenine riboswitches. (a)
For the \emph{add} adenine riboswitch aptamer the secondary structure is on the left, and the tertiary structure is on the right. 
The tertiary structure for {\it add} A-riboswitch aptamer is taken from the
crystal structure (PDB Id: 1Y26).
(b) Same as (a) except the structures corresponds to {\it pbuE} adenine
riboswitch aptamer. The tertiary structure, corresponding to a
snapshot at $t = 95$ ns in the molecular dynamics simulation, is merely a model.}
\end{center}
\end{figure}

\begin{figure}{}
\begin{center}
\centerline{\includegraphics[width=.75\textwidth]{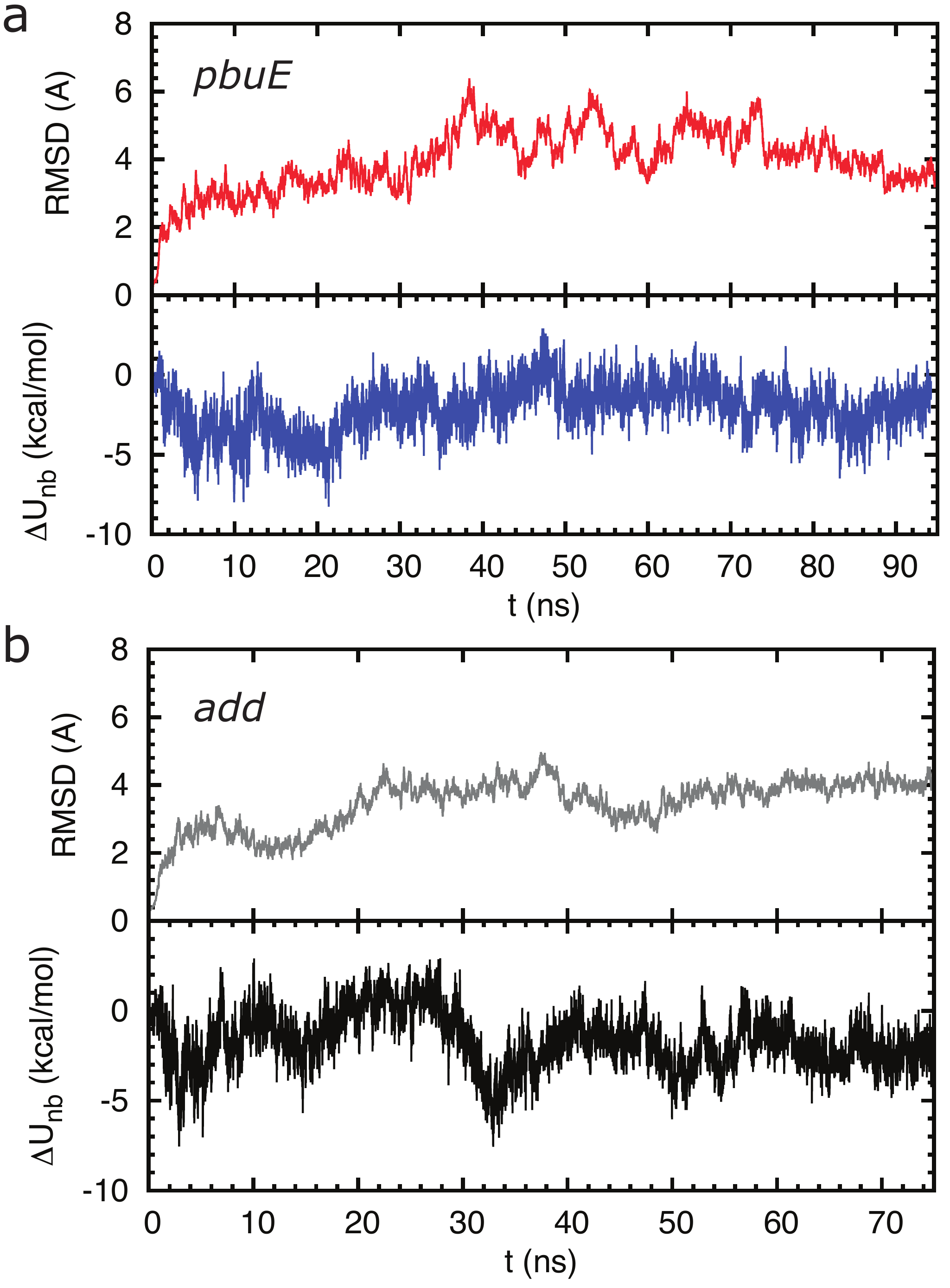}}
\caption{(a) Stability of the folded {\it pbuE} A-riboswitch aptamers
inferred from all atom MD simulation. (Upper panel) Time
evolution of the root mean square deviation (RMSD) of backbone nucleotide
positions. (Lower panel) Fluctuations in
the difference between non-bonded energies of helices P2 and P3,
$\Delta U_{nb} = U_{nb,P2} - U_{nb,P3}$, in the aptamer with the metabolite
bound during the all atom molecular dynamic simulations.
The curve suggests that the putative MD structures for \emph{pbuE} 
 A-riboswitches are stable. 
(b) Same as (a) except for the {\it add} A-riboswitch aptamer. Here, $\Delta U_{nb} = U_{nb,P3} - U_{nb,P2}$}
\end{center}
\end{figure}

\begin{figure}{}
\begin{center}
\centerline{\includegraphics[width=.75\textwidth]{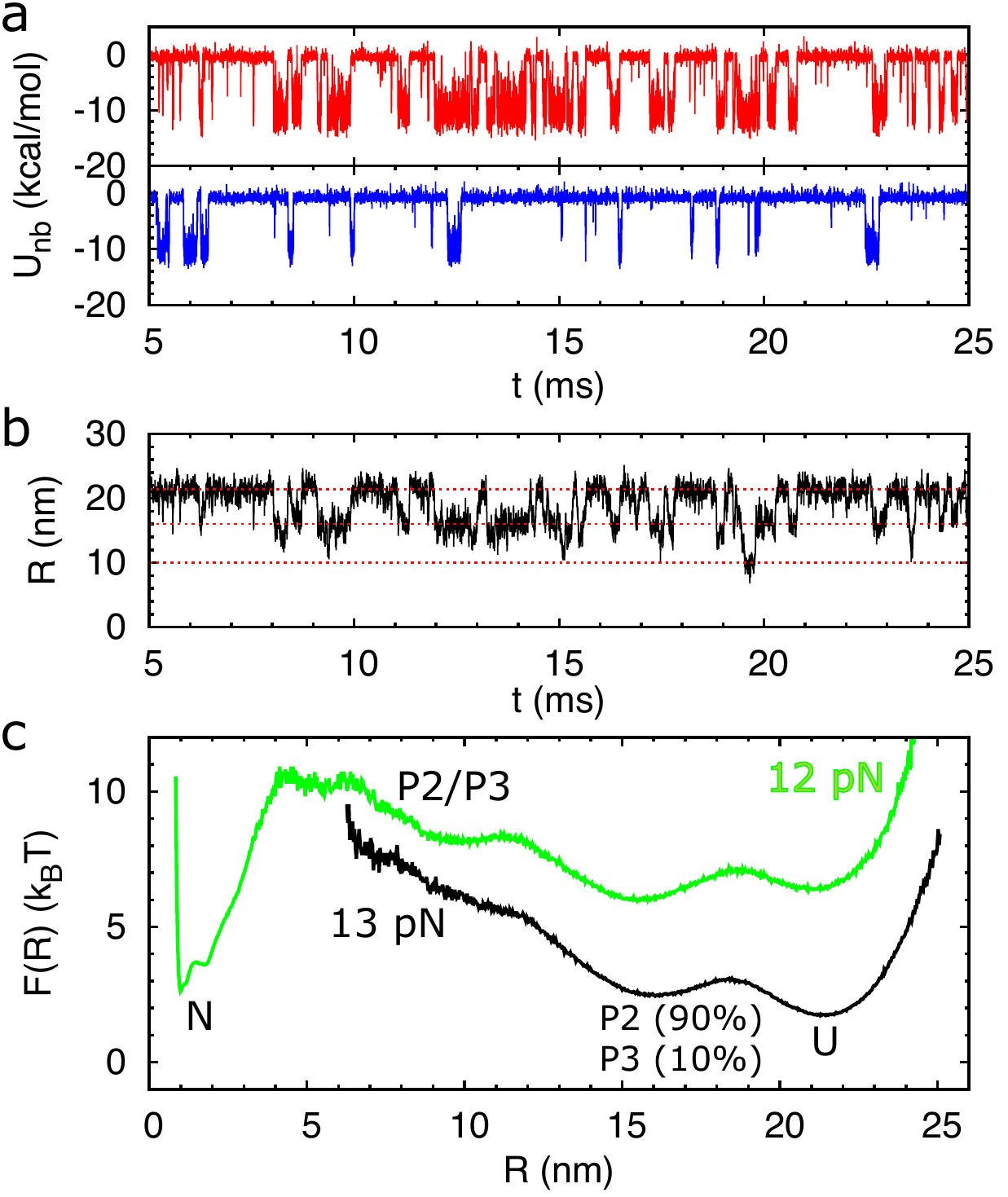}}
\caption{Coarse-grained pulling simulations for \emph{pbuE} A-riboswitch
using MD generated structures. 
(a) The time traces of non-bonded energies of helices P2 (red line) and 
P3 (blue line) in a trajectory showing that P2 spends more time
in the folded state than P3
for the {\it pbuE} A-riboswitch aptamer at $f = 13$ pN.
(b) The time traces of the 
end-to-end distance, $R$, of the riboswitch for the trajectory in (a).
(c) The free energy profile obtained based on the time traces for $f=12$ (green line) and 13 pN (black line). 
Interestingly, when $f$ is increased by 1 pN from $f=12$ to $f=13$ pN the folded state is completely destabilized. }
\end{center}
\end{figure}

\begin{figure}{}
\begin{center}
\centerline{\includegraphics[width=.75\textwidth]{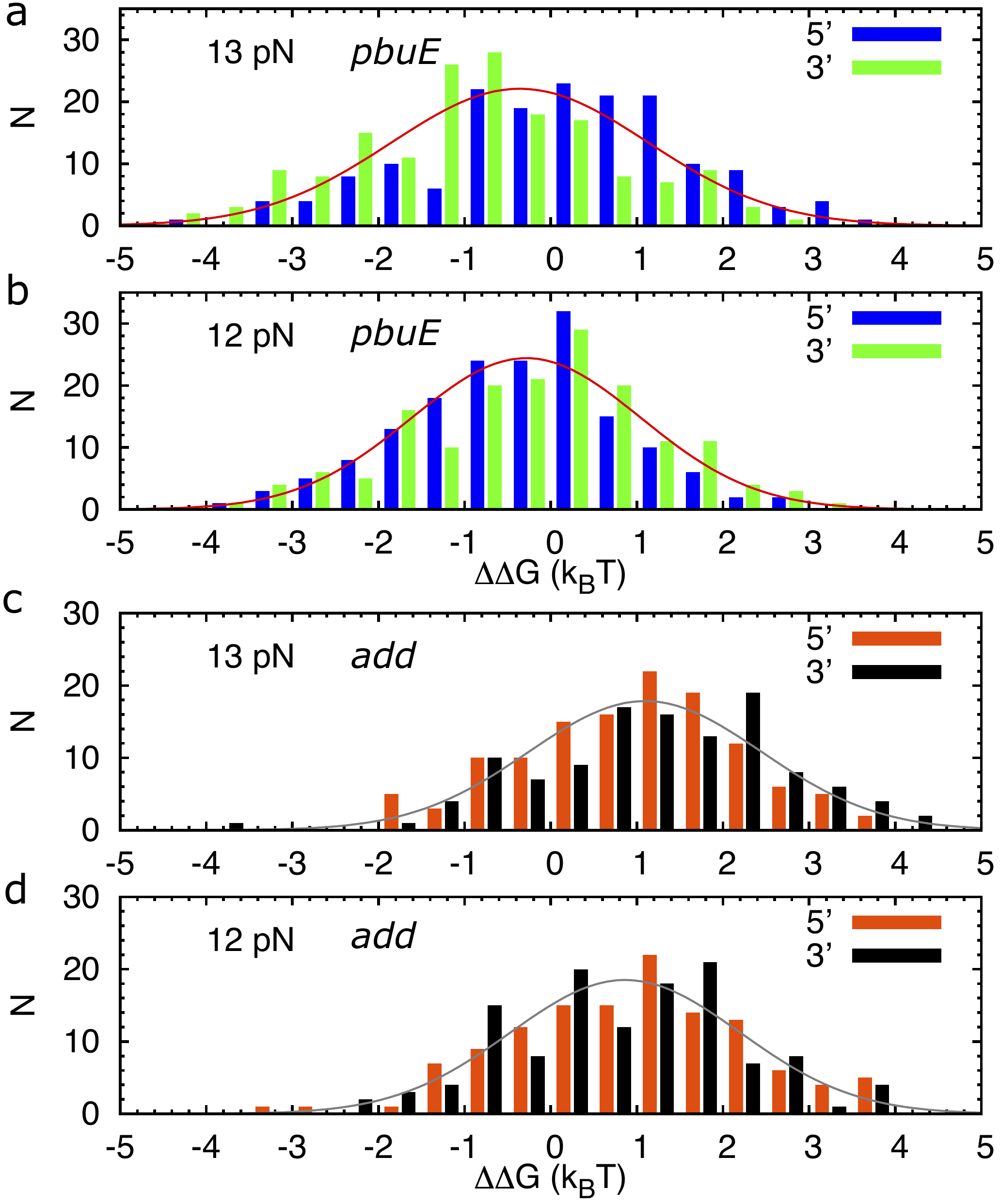}}
\caption{The histograms of the difference in the free energy difference
between only-P2-folded and only-P3-folded states,
{\it i.e.}, $\Delta \Delta G = - k_{B}T$ln$(F_{P2}/F_{P3})$, for (a) total 168
structures for the {\it pbuE} A-riboswitch aptamer with forces, $f = 13$ pN
and (b) $f = 12$ pN, applied on either end of the aptamer in the coarse-grained
simulations.
(c) Same as (a) except for total 131 trajectories for
the {\it add} A-riboswitch aptamer with forces, $f = 13$ pN and (d) $f = 12$ pN.}
\end{center}
\end{figure}

\end{document}